\documentstyle[twocolumn,prl,aps,epsf,epsfig,amsmath]{revtex}  
  
\font\cmss=cmss12   
\def\1{\hbox{{1}\kern-.25em\hbox{l}}}  
\def\bfZ{\relax{\hbox{\cmss Z\kern-.4em Z}}}  
  
\newcommand{\pz}{\bar P \cdot z}  
\newcommand{\rz}{r \cdot z}  
\newcommand{\beq}{\begin{equation}}  
\newcommand{\eeq}{\end{equation}}  
\newcommand{\bea}{\begin{eqnarray}}  
\newcommand{\eea}{\end{eqnarray}}  
 
\begin{document}  
  
\title{Next-to-leading order evolution of generalized parton distributions for DESY HERA and HERMES}  
\author{Andreas Freund\thanks{andreas.freund@physik.uni-regensburg.de}}
\address{Institut f{\"u}r Theoretische Physik, University of Regensburg, 
Universit{\"a}tstr. 31, 93053 Regensburg, Germany}
\author{Martin McDermott\thanks{martinmc@amtp.liv.ac.uk}}  
\address{Division of Theoretical Physics, Dept. Math. Sciences, University of Liverpool, Liverpool, L69 3BX, UK}  
\maketitle  
 
\begin{abstract}  
The QCD evolution of both unpolarized and polarized generalized parton distributions (GPDs) to next-to-leading order (NLO) accuracy is presented, in both the DGLAP and ERBL regions, for two appropriately symmetrized input distributions based on conventional parton density functions. To illustrate the relative size of the NLO corrections a comparison is made with leading order evolution of the same distributions. For the first time, NLO results are given for both small and large values of the skewedness parameter, $\zeta = x_{bj}$, i.e. for all of the kinematic range relevant to HERA and HERMES.  
\vspace{1pc}  
\end{abstract}  
 
\section{Introduction}  
  
Generalized Partons Distributions (GPDs) \cite{mrgdh,ji,rad1,cfs} are 
generic two-parton correlation functions of nucleons that provide  
the boundary conditions for the calculation of various hard, exclusive, diffractive processes in QCD.  
For such processes, which are characterised by a suitable hard scale  
(for example a large photon virtuality, $Q^2$, in Deep Inelastic Scattering (DIS) experiments),  
the incoming nucleon remains intact and at high energies is well separated in rapidity  
from the rest of the final state, i.\ e.\ the diffractively produced 
low mass particle (e.g. a meson or real photon). A perturbative QCD analysis is possible provided that a  
factorization theorem is proved which illustrates the separation of short and long distance physics.  
The QCD amplitude is then a sum of terms involving a convolution of a hard scattering coefficient  
(calculated to a given order in $\alpha_s$) with a GPD (specified to a related accuracy in logarithms) and possibly a second convolution with a distribution amplitude, in the case of the production of a hadronic final state.  

The universal, non-perturbative GPDs obey renormalization group equations (RGE)  
and are formally defined, up to power suppressed contributions, by Fourier transforms of  
non-local, renormalized, light-cone operators sandwiched between nucleon states of {\it unequal momentum}  
(the corresponding distributions in inclusive DIS involve equal momenta for the incoming and  
outgoing nucleon). In common with conventional DGLAP parton distribution functions,  
GPDs cannot be calculated from first principles at present due to the confinement problem,  
but they can be pinned down through a global RGE analysis of all available experimental data.  
The simplest example of such a hard, exclusive, diffractive process is deeply virtual Compton  
scattering (DVCS) \cite{mrgdh,ji,cf,jio}, for which there is recent first data available from  
HERA \cite{zeus,h1}, HERMES \cite{hermes} and JLAB \cite{jlab}.  
  
The RGE for the GPDs involve kernels which have been computed up to next-to-leading order (NLO)  
in perturbation theory so far \cite{bfm}. The solutions of these NLO evolution equations are necessary  
to obtain accurate predictions for various experimentally accessible processes.  
In fact the DVCS amplitude itself can be directly accessed experimentally through the  
interference of DVCS with the Bethe-Heitler process \cite{zeus,h1,hermes}.  
In this paper we present numerical solutions to the NLO QCD evolution of the GPDs for large (HERMES) and small (HERA) skewedness which corresponds to large and small 
Bjorken $x$ in the particular representation used.  
As input for the NLO evolution, we use a model based on double 
distributions (DDs) \cite{rad2,musrad} incorporating for comparison  
two related sets of conventional NLO unpolarized and polarized parton 
distribution functions (PDFs), MRSA$^{'}$ \cite{mrsap} and 
 Gehrmann-Stirling (`GS(A)')\cite{gehrst}, and GRV98 \cite{grv98} and GRSV00 
\cite{grsv00}, respectively. The resultant input GPDs,  
by design, respect all the known  
symmetries and properties of the GPDs, which are preserved under evolution. This latter condition constitutes a  
stringent test of any proposed solutions, which is passed by our numerical solutions to very good accuracy.  
  
This paper presents the first results of the numerical solution to the RGEs at NLO accuracy \cite{foot1,bmns1,bmns2} for any $x$ (see, e.g., \cite{vgaf,kgbm} for leading order skewed evolution) and is organised as follows.  
In section \ref{sec:def} we give the formal definitions of the GPDs and discuss their symmetry and spectral properties.  
Section \ref{sec:inp} describes our choice of input distributions to 
the evolution equations outlined in section  
\ref{sec:evol}.  
We present our numerical results, including a comparison between leading order and next-to-leading order skewed evolution and a comparison with conventional next-to-leading order evolution, in section \ref{sec:res}, and conclude in section \ref{sec:con}. Our 
evolution code will be made available to the community in the  
near future via the internet \cite{website}. 
  
\section{Definitions, symmetries and spectral representations of GPDs}  
  
\label{sec:def}  
  
Amplitudes for quark and gluon correlators of unequal momentum nucleon 
states may be defined in a number of ways. We choose a definition which treats  
the initial and final state nucleon momentum ($p,p'$, respectively) symmetrically by involving parton light-cone fractions with respect to the momentum transfer, $r = 
p - p'$, and the average momentum, $\bar P = (p + p')/2$.  
The flavour singlet and non-singlet (S,NS) quark, and the gluon (G) matrix elements of the non-local operators, involving a light-cone vector $z^{\mu}$ ($z^2=0$), are defined by  
\bea  
&&2 M_a^{NS}(\pz,\rz,t) = \nonumber\\  
&&\langle N\left(P_{-}\right)|{\bar \psi}_a\left(-\frac{z}{2}\right){\cal 
P}\hat z\psi_a\left(\frac{z}{2}\right)|N\left(P_{+}\right)\rangle 
+\nonumber\\ 
&&\langle N\left(P_{-}\right)|{\bar \psi}_a\left(\frac{z}{2}\right){\cal 
P}\hat z\psi_a\left(-\frac{z}{2}\right)|N\left(P_{+}\right)\rangle \, , \nonumber\\  
&&2 M^{S}(\pz,\rz,t) = \nonumber\\  
&&\sum_{a}\langle N\left(P_{-}\right)|{\bar 
\psi}_a\left(-\frac{z}{2}\right){\cal P}\hat z\psi_a\left(\frac{z}{2}\right)|N\left(P_{+}\right)\rangle - \nonumber\\  
&&\langle N\left(P_{-}\right)|{\bar \psi}_a\left(\frac{z}{2}\right){\cal P}\hat z\psi_a\left(-\frac{z}{2}\right)|N\left(P_{+}\right)\rangle \, , \nonumber\\ 
&&2 M^G(\pz,\rz,t) = \nonumber\\  
&&z^{\mu}z^{\nu}\langle N\left(P_{-}\right)|G_{\mu\rho}\left(-\frac{z}{2}\right){\cal 
P}G^{\rho}_{\nu}\left(\frac{z}{2}\right)|N\left(P_{+}\right)\rangle +\nonumber\\ 
&&z^{\mu}z^{\nu}\langle N\left(P_{-}\right)|G_{\mu\rho}\left(\frac{z}{2}\right){\cal 
P}G^{\rho}_{\nu}\left(-\frac{z}{2}\right)|N\left(P_{+}\right)\rangle ,\nonumber\\ 
\eea 
where $P_{+,-} = \bar P \pm \frac{r}{2}$, $t=r^2$ is the four-momentum 
transfer, ${\hat z} = \gamma^{\mu}z_{\mu}$, $\pz$ and $\rz$ are dimensionless 
Lorentz scalars, $a = u,d,s,c...$ is a flavour index and the symbol $\cal P$ 
represents the usual path ordered exponential. The spin and isospin dependence 
have been suppressed for convenience.  
Polarized matrix elements are defined in a similar fashion, but with  
$\hat z \to \hat z \gamma^5$ for the quark and with the implicit metric  
$g^{\mu\nu} \to i\epsilon^{\mu\nu-+}$ for the gluon.  
The unpolarized matrix elements, and thus their associated GPDs, have 
definite symmetry properties. G-parity gives (suppressing the $t$-dependence)  
\begin{align} 
&M^{NS}(\pz,\rz)=-M^{NS}(-\pz,-\rz) \nonumber\\  
&M^{S,G}(\pz,\rz)= M^{S,G}(-\pz,-\rz) \, .  
\label{sym}  
\end{align} 
\noindent The opposite properties are observed for the polarized case.  
Hermitian conjugation gives for all polarized and unpolarized species ($i= S,NS,G$)  
\bea 
M^{i}(\pz,\rz) = M^{i}(\pz,-\rz) \, .  
\label{herm}  
\eea 
 
The matrix elements can be most generally represented by a double spectral  
representation with respect to $\pz$ and $\rz$ \cite{mrgdh,rad1,pw} as follows:  
\begin{align} 
&M^{NS}=\int^1_{-1}dx e^{-ix\pz}\int^{1-|x|}_{-1+|x|}dy e^{-iy\rz/2} \times \, \, \, \nonumber\\  
&\left[\bar U'\hat z U F^{NS}(x,y,t) + \frac{iz^{\mu}r^{\nu}\bar U' 
\sigma_{\mu\nu} U}{2m_N} K^{NS}(x,y,t)\right] \, ,  
\nonumber\\  
&M^{G} = \int^1_{-1}dx e^{-ix\pz} \int^{1-|x|}_{-1+|x|}dy e^{-iy\rz/2} \times \nonumber\\  
&\Big[ \bar U'\hat z U \frac{\pz}{2} F^G(x,y,t) \, \nonumber\\ 
&+ \frac{iz^{\mu}r^{\nu}\bar U' \sigma_{\mu\nu} U \rz}{4m_N} K^G(x,y,t) \Big] \nonumber\\ 
&+ {\bar U'} U\rz\int^1_{-1}dy e^{-iy\rz/2} D^G (y,t) \, ,  \nonumber\\  
&M^{S}=\int^1_{-1}dx e^{-ix\pz}\int^{1-|x|}_{-1+|x|}dy e^{-iy\rz/2} \times \nonumber\\  
&\left[\bar U'\hat z U F^{S}(x,y,t) + \frac{iz^{\mu}r^{\nu}\bar U' \sigma_{\mu\nu} U}{2m_N} K^{S}(x,y,t) \right]\nonumber\\&+ {\bar U'} U\rz\int^1_{-1}dy e^{-iy\rz/2} D (y,t) \, , \label{dd}  
\end{align}  
\noindent where $\bar U^{'}$ and $U$ are nucleon spinors. Note that in accordance with  
the associated Lorentz structures, the $F$'s correspond to helicity non-flip  
and the $K$'s to helicity flip amplitudes, and are collectively known as double distributions.  
Henceforth, for brevity, we shall only discuss the  
helicity non-flip piece explicitly. However, the helicity flip case is exactly analogous.  
The $D$-terms in the last two lines permit non-zero values for the singlet $M^{S}$ and $M^{G}$ 
in the limit $\pz \to 0$ and $\rz \neq 0$, which is  
allowed by their evenness in $\pz$ (cf. eq.(\ref{sym})).  
Conversely, $M^{NS}$ is required by its oddness under 
$\pz$, for any $\rz$, to be zero \cite{pw}.  
  
By making a particular choice of the light-cone vector, $z^{\mu}$, as a light-ray vector  
(so that in light-cone variables, $z_{\pm} = z_0 \pm z_3 $, only its minus component is  
non-zero $z^{\mu} = (0,z_{-},0)$) one may reduce the double spectral representation  
of eq.(\ref{dd}), defined on the entire light-cone, to a one 
dimensional spectral representation, defined along a light ray, depending on the skewedness  
parameter, $\xi$, defined by  
\begin{align} 
\xi = \rz /2 \pz = r_{+}/ 2 {\bar P}_{+} \, . 
\end{align} 
The resultant GPDs are the off-forward distribution functions (OFPDFs) introduced in \cite{mrgdh,ji}: 
\begin{align} 
&H(v,\xi,t) = \nonumber\\  
&\int^1_{-1}dx' \int^{1-|x'|}_{-1+|x'|}dy' \delta(x'+ \xi y' - v) F(x',y',t) \, ,  
\label{1dim}  
\end{align}  
\noindent where $v \in [-1,1]$. In terms of individual flavour decomposition the singlet, non-singlet and gluon distributions are given through  
\begin{align} 
&H^S(v,\xi) = \sum_a H^{q,a} (v,\xi) \mp H^{q,a} (-v,\xi) \, ,\nonumber\\  
&H^{NS,a}(v,\xi) = H^{q,a} (v,\xi) \pm H^{q,a} (-v,\xi)   \, , \nonumber\\  
&H^G(v,\xi) = H^g(v,\xi) \pm H^g(-v,\xi) \, , 
\end{align} 
where the upper (lower) signs corresponds to the unpolarized (polarized) case.  
The unpolarized singlet and the gluon have additional resonance-like 
contributions called the D-term \cite{pw,dterm} which ensures the correct 
symmetry and polynomiality properties (cf. eq.(\ref{dd})).  
Note that the symmetries which hold for the matrix elements change for the $H^i$s,  
due to the $\pz$, $\hat z$ factors in eq.(\ref{dd}). The unpolarized quark singlet is  
antisymmetric about $v=0$, whereas the unpolarized quark non-singlet and the gluon are  
symmetric. The opposite symmetries hold for the polarized distributions.  
The helicity flip GPDs, (labelled with $E$s), are found analogously (double integrals with respect to the $K$s).  
  
Henceforth, we shall make the usual assumption that the $t$-dependence of all of these  
functions factorizes into implicit form factors. One should bear in mind that in order  
to make predictions for physical amplitudes (for $t \neq 0$) these form factors must be specified.  
Note that the assumption of a factorized $t$-dependence, as a general statement, must be justified within the kinematic regime concerned. 
It appears to be valid at small $x$ and small $t$, from the HERA data on a variety of diffractive measurements. However, it appears to be wrong for moderate to large $t$ and larger $x$ \cite{ppg}.   
The evolution used in this note assumes the same factorized $t$-dependence for 
each parton species, which therefore factorizes in the evolution equations. Thus any effect  
of assuming a different $t$-dependence for quark singlet and gluon, which mix under evolution,  
is neglected (but could in principle be investigated within the same framework). 
The effect on evolution of relaxing the assumption of factorization of the $t$-dependence altogether remains an open question. 
 
Taking the $N-1$ moments (for even $N$) of the GPDs yields a sum of polynomials  
in the skewedness parameter $\xi$ \cite{pw}:  
\begin{align}  
&\int^1_{-1}dv v^{N-1}H(v,\xi,t) =\nonumber\\  
&\sum^{N}_{k=0}\xi^k\left(N\atop k\right)  
\int^1_{-1}dx\int^{1-|x|}_{-1+|x|}F(x,y,t)x^{N-k}y^k \, . 
\label{poly}  
\end{align} 
This powerful polynomiality condition arises because $\mu_i$ indices of  
a local operator $O^{\mu_1...\mu_N}$ must be carried by either ${\bar P}$ or $r$  
(taking moments is equivalent to going from non-local to local operators).  
In building models for GPDs, eq.(\ref{poly}) always has to be fulfilled and  
thus strongly constrains any model for the GPDs. We incorporate it in a natural way  
in building our input distributions in section \ref{sec:inp}.  
  
For the purposes of comparing to experiment it is natural to define GPDs in terms of  
momentum fractions, $X \in [0,1]$, of the incoming proton momentum, $p$, carried by the outgoing  
parton. To this end we adapt the notation and definitions of \cite{kgbm} introducing two non-diagonal parton distribution functions (NDPDFs), ${\cal F}^q$ and ${\cal F}^{\bar q}$, for flavour, $a$:  
\begin{align} 
&{\cal F}^{q,a} \left (X_1=\frac{v_1 + \xi}{1+\xi},\zeta\right ) = \frac{H^{q,a} (v_1,\xi)}{(1-\zeta/2)} \, , \label{fq} \\  
&{\cal F}^{{\bar q},a} \left (X_2=\frac{\xi - v_2}{1+\xi},\zeta\right ) = -\frac{H^{q,a} (v_2,\xi)}{(1-\zeta/2)} \, , \label{fqbar}  
\end{align}  
where $v_1 \in [-\xi,1], \, v_2 \in [-1,\xi]$ (see fig.(4) of \cite{kgbm}),  
$\zeta \equiv r_+/p_+ $ is the skewedness defined on the domain $ \zeta \in [0,1]$ such that  
$\xi = \zeta/(2-\zeta)$ and $\zeta = x_{bj}$ for DVCS, up to terms of 
${\cal O} (x_{bj} t / Q^2)$.  
The inverse transformations between the $v$s and $X$s is:  
\bea  
v_1 = \frac{X_1 - \zeta/2}{1-\zeta/2} & \, , \, & v_2 = \frac{\zeta/2 - X_2}{1-\zeta/2} \, . 
\eea  
For the gluon one may use either transformation, e.g.  
\bea  
{\cal F}^{g} (X,\zeta) &=& \frac{H^g (v_1, \xi)}{(1-\zeta/2)} \, . 
\eea  
    
There are two distinct kinematic regions for the GPDs, with different physical interpretations.  
In the DGLAP \cite{dglap} region, $X > \zeta$ ($|v| > \xi$), ${\cal F}^q (X,\zeta)$ and  
${\cal F}^{\bar q}(X,\zeta)$ are independent functions, corresponding to quark or anti-quark  
fields leaving the nucleon with momentum fraction $X$ and returning with positive momentum  
fraction $X-\zeta$. As such they correspond to a generalization of regular DGLAP PDFs  
(which have equal outgoing and returning fractions). In the ERBL \cite{erbl} region,  
$X<\zeta$ ($|v| < \xi$), both quark and anti-quark carry positive momentum  
fractions ($X,\zeta-X$) {\it away} from the nucleon in a meson-like configuration, and the  
GPDs behave like ERBL \cite{erbl} distributional amplitudes characterising mesons.  
This implies that ${\cal F}^q $ and ${\cal F}^{\bar q}$ are not independent in the ERBL region, indeed a symmetry is observed: ${\cal F}^q (\zeta-X,\zeta) = {\cal F}^{\bar q} (X,\zeta)$ (which directly reflects the symmetry of  
$H^q(v,\xi)$ about $v=0$). Similarly, the gluon distribution, ${\cal F}^g$, is DGLAP-like for  
$X>\zeta$ and ERBL-like for $X<\zeta$. This leads to unpolarized non-singlet,  
${\cal F}^{NS,a} = {\cal F}^{q,a} - {\cal F}^{{\bar q},a}$,  and gluon GPDs which are symmetric,  
and a singlet quark distribution ${\cal F}^S = \sum_a {\cal F}^{q,a} + {\cal F}^{{\bar q},a}$  
which is antisymmetric, about the point $X = \zeta/2$ in the ERBL region.  
Again the opposite symmetries hold for the polarized distributions.

\section{Input GPDs}  
  
\label{sec:inp}  
  
In this section we describe how to build input distributions,  
${\cal F}^{q, {\bar q}, g} (X,\zeta, Q_0) $, at the input scale, $Q_0$, with the correct symmetries  
and properties from conventional PDFs in the DGLAP region, for both the unpolarized and polarized  
cases. These input NDPDFs then serve as the boundary conditions for our numerical evolution.  
  
Factoring out the overall $t$-dependence we have for the quark for example ($v_1 > -\xi$):  
\begin{align}  
&{\cal F}^{q,a} (X,\zeta) = \frac{H^{q,a} (v_1,\xi)}{1-\zeta/2} = \nonumber\\  
&\int^{1}_{-1} dx' \int^{1-|x'|}_{-1+|x'|} dy' \delta \left( x' + \xi y' - v_1 \right)  
\frac{F^{q,a} (x',y')}{\left(1-\zeta/2\right)} \, .  
\label{fqinp}  
\end{align}  
  
Following \cite{rad2,musrad,bmns2} we employ a factorized ansatz which expresses the double distribution  
as a product of a profile function, $\pi^{i}$, and a conventional PDF, $f^{i}$, $(i=q,g$):  
\bea 
F^{q,a}(x',y') &=& \pi^{q} (x',y') f^{q,a} (x') \nonumber\\  
    &=& \frac{3}{4} \frac{(1-|x'|)^2 - {y'}^2}{(1-|x'|)^3} f^{q,a} (x') \, , \nonumber\\  
F^g(x',y')  &=& \pi^g(x',y') f^g (x') \nonumber\\  
    &=& \frac{15}{16} \frac{((1-|x'|)^2 - {y'}^2)^2}{(1-|x'|)^5} f^g (x') \, , \label{ddinp} 
\eea 
\noindent for quark of flavour $a$, where  
\bea  
f^g (x)  &=& xg(x,Q_0) \Theta(x) + |x| g (|x|,Q_0) \Theta (-x) \, , \nonumber \\  
f^{q,a} (x) &=& q^{a} (x,Q_0) \Theta(x) - (\bar q^{a}) (|x|,Q_0) \Theta (-x) \, . \label{pdfinp} 
\eea  
The profile functions are chosen to guarantee the correct symmetry properties  
in the ERBL region and their normalization is specified by demanding that the  
conventional distributions are reproduced in the forward limit: e.g.    
${\cal F}^{g} (X,\zeta \to 0) \to f^g (X)$.   
We have explicitly checked this forward limit for our input codes against independent  
codes for the conventional PDFs (see fig.(\ref{fig:ratfwd}) later).  
  
So far, models for polarized distributions have been strongly based,   
for reasons of consistency, on closely associated models for unpolarized  
distributions. For this reason we choose as our inputs these closely related pairs  
of models. Unfortunately, the polarized set of distributions of Gehrmann and  
Stirling (`GS(A)') \cite{gehrst} is rather old now and is based on the MRSA$^{'}$ unpolarized set \cite{mrsap}.  
Our other set is the very recent GRSV set from 2000 \cite{grsv00} which is based on the 1998  
unpolarized model by GRV \cite{grv98} (we will denote these GRSV00 and GRV98 respectively  
\cite{foot2}).  
  
Having defined this model for the double distribution one may then perform the $y'$-integration in eq.(\ref{fqinp}) using the delta function. This modifies the limits on  
the $x'$ integration according to the region concerned: for the DGLAP region $X> \zeta$ one has:  
\begin{align}  
&{\cal F}^{q,a} (X,\zeta) = \nonumber\\ 
&\frac{2}{\zeta} \int^{\frac{v_1+\xi}{1+\xi}}_{\frac{v_1-\xi}{1-\xi}} dx'  
\pi^q \left (x', \frac{v_1 - x'}{\xi} \right) q^a (x') \, . 
\end{align}  
For the anti-quark, since $v_2 = -v_1$ one may use eqs.(\ref{fqbar},\ref{fqinp}) with $v_2, v_1 \to -v_1$,  
and exploiting the fact that $f^{q} (x) = - {\bar q} (|x|) $ for $ x < 0 $, one arrives at  
\begin{align}  
&{\cal F}^{\bar q,a} (X,\zeta) = \nonumber\\ 
&\frac{2}{\zeta} 
\int^{\frac{-v_1+\xi}{1-\xi}}_{\frac{-v_1-\xi}{1+\xi}} dx' \pi^q \left  
(x',\frac{-v_1 - x'}{\xi} \right ){\bar q}^a(|x'|).  
\end{align} 
The non-singlet (valence) and singlet quark combinations are given by:  
\bea  
{\cal F}^{NS,a} & \equiv & {\cal F}^{q,a} - {\cal F}^{{\bar q},a} \nonumber \\  
                & \equiv & \frac{[H^{q,a} (v_1,\xi) + H^{q,a} (-v_1,\xi)]}{1-\zeta/2} \, , \label{fnsing} \nonumber \\  
{\cal F}^{S} & \equiv & \sum_a {\cal F}^{q,a} + {\cal F}^{{\bar q},a} \nonumber \\  
                & \equiv & \sum_a \frac{[H^{q,a} (v_1,\xi) - H^{q,a} (-v_1,\xi)]}{1-\zeta/2} \, . \label{fsing}  
\eea  
  
In the ERBL region ($X>\zeta, |v| < \xi$) integration over $y'$ leads to:  
\begin{align} 
&{\cal F}^{q,a} (X,\zeta) = \frac{2}{\zeta} \times \nonumber \\  
&\left[ \int^{\frac{v_1+\xi}{1+\xi}}_{0} dx' \pi^q \left(x', \frac{v_1 - x'}{\xi} \right) q^a (x') \, - \right. \nonumber \\  
&\left. \int^{0}_{\frac{-(\xi-v_1)}{1+\xi}} dx' \pi^q \left(x', \frac{v_1 - 
x'}{\xi} \right) {\bar q}^a (|x'|) \right] \, ,  
\nonumber \\  
&{\cal F}^{{\bar q},a} (X,\zeta) = -\frac{2}{\zeta} \times \nonumber \\  
&\left[ \int^{\frac{\xi-v_1}{1+\xi}}_{0} dx' \pi^q \left(x', \frac{-v_1 - x'}{\xi} \right) q^a (x') \, - \right. \nonumber \\  
&\left. \int^{0}_{\frac{-(\xi+v_1)}{1+\xi}} dx' \pi^q \left(x', \frac{-v_1 - x'}{\xi} \right) {\bar q}^a (|x'|) \right] \, .  
\end{align} 
This gives singlet and non-singlet distributions in the ERBL region containing four integrals over $x'$.  
The unpolarized singlet also includes the D-term on the right hand side of eq.(\ref{fsing})  
(in principle there is also an analogous term in the unpolarized gluon
($D^G$ in eq.(\ref{dd})), but we choose to set this to zero, since 
nothing is known for the gluon D-term except its symmetry.). We adapt 
the model introduced in \cite{dterm} for the unpolarized singlet D-term 
based on the chiral-soliton model:  
\beq  
{\cal F}^{D} (X,\zeta) = \Theta(\zeta -X) \, D\left(\frac{2X}{\zeta} -1, t=0\right) \, , 
\end{equation}
\noindent with $D$ given by a truncated expansion in terms of odd Gegenbauer Polynomials:  
\begin{align}  
&D(a) = (1-a^2) \times  \nonumber \\  
&\left[ -4.0 C_1^{3/2} (a) - 1.2 C_3^{3/2} (a) - 0.4 C_5^{3/2} (a) \right] \, .  
\label{eq:dterm} 
\end{align} 
This D-term is antisymmetric in its argument, i.\ e.\ about the point $X = \zeta/2$ (in keeping with the anti-symmetry of ${\cal F}^S$, and $H^S$ about $v=0$). It vanishes entirely in the forward limit and only assumes numerical significance for large $\zeta$.  
This completes the definition of the input singlet and non-singlet quark distributions in both regions. The input gluon is  
defined along the same lines as the non-singlet.  
  
\section{NLO evolution of GPDs}  
  
\label{sec:evol}  
  
The input GPDs, defined above, are continuous functions which span the
DGLAP and ERBL regions, and evolve in scale appropriately according to
the DGLAP or ERBL evolution equations, given below. Note that the
evolution in the ERBL region depends on the DGLAP region (see eq.\
(\ref{erbleq})) whereas the DGLAP evolution is independent of the ERBL 
region. One remarkable feature is that the evolved functions remain
continuous under evolution and continue to satisfy all appropriate
symmetries in the ERBL region. As can be seen, our numerical solutions  
manifestly exhibit these properties (see Section \ref{sec:res}), thus
making us confident in the correctness of our numerical solutions.  
  
In the DGLAP region the singlet and gluon distributions mix under evolution:  
\begin{align}  
&\frac{d {\cal F}^S (y,\zeta,Q^2)}{d\ln(Q^2)} =  
\int^1_y \frac{dz}{z} P_{qq}\left(\frac{y}{z},\frac{\zeta}{z}\right)_+{\cal F}^S (z,\zeta,Q^2)  
\nonumber\\  
&+ \left(1-\frac{\zeta}{2}\right)\int^1_y \frac{dz}{z} P_{qg}\left(\frac{y}{z},\frac{\zeta}{z}  
\right){\cal F}^G(z,\zeta,Q^2) \, , \nonumber\\  
&\frac{d {\cal F}^G(y,\zeta,Q^2)}{d\ln(Q^2)} =  
\int^1_y \frac{dz}{z} P_{gg}\left(\frac{y}{z},\frac{\zeta}{z}\right)_+ {\cal F} ^G(z,\zeta,Q^2)  
\nonumber\\  
&+\frac{1}{1-\frac{\zeta}{2}}\int^1_y \frac{dz}{z} P_{gq} \left(\frac{y}{z},\frac{\zeta}{z}  
\right) {\cal F}^S (z,\zeta,Q^2) \, , \label{dglapeq}   
\end{align} 
\noindent with generalized DGLAP kernels \cite{bfm}. The NS combinations do not mix under  
evolution so we omit them for brevity.  
  
The $+$-distribution for the DGLAP kernel, regulates the divergence as 
$z \to y$. In general it is defined as follows:
\begin{equation}
P(x,\zeta)_+ = P(x,\zeta) - \delta(x-1) \int^{1}_{0} dx' P(x',\zeta)
\end{equation}
The numerical implementation of the $+$-distribution applied to the integrals in eq.(\ref{dglapeq}) is as follows:
\begin{align}  
&\int^1_y \frac{dz}{z} P\left(\frac{y}{z},\frac{\zeta}{z}\right)_+{\cal F}(z,\zeta) = \nonumber\\  
&\int^1_y \frac{dz}{z} P\left(\frac{y}{z},\frac{\zeta}{z}\right)\left( {\cal F} (z,\zeta) - {\cal F}(y,\zeta)\right) - \nonumber\\  
&{\cal F}(y,\zeta)\Big[\int^1_{\frac{\zeta}{y}} dz P\left(z,\frac{\zeta}{y}\right)  
-\int^1_y \frac{dz}{z} P\left(z,z\frac{\zeta}{y}\right)\nonumber\\
&+\frac{y}{\zeta}\int^{\zeta/y}_0 dz V\left(z\frac{y}{\zeta},\frac{y}{\zeta}\right)\Big]\, .  
\end{align} 
  
In the ERBL region we have  
\begin{align}  
&\frac{d {\cal F}^S(y,\zeta,Q^2)}{d\ln(Q^2)}= \Big[\int^1_y \frac{dz}{\zeta} V^{qq}\left(\frac{y}{\zeta},  
\frac{z}{\zeta}\right)_+ +\nonumber\\  
&\int^y_0 \frac{dz}{\zeta} V^{qq}\left(\overline {\frac{y}{\zeta}},\overline {\frac{z}{\zeta}}\right)_+  
\mp \int^1_{\zeta} \frac{dz}{\zeta} V^{qq}\left(\overline{\frac{y}{\zeta}},\frac{z}{\zeta}\right)\Big]  
{\cal F}^S\nonumber\\  
&+\left(1-\frac{\zeta}{2}\right)\Big[\int^1_y \frac{dz}{\zeta^2}  
V^{qg}\left(\frac{y}{\zeta},\frac{z}{\zeta}\right)\nonumber\\  
&-\int^y_0 \frac{dz}{\zeta^2} V^{qg}\left(\overline {\frac{y}{\zeta}},\overline {\frac{z}{\zeta}}\right)  
\mp \int^1_{\zeta} \frac{dz}{\zeta^2}  
V^{qg}\left(\overline{\frac{y}{\zeta}},\frac{z}{\zeta}\right)\Big] {\cal F}^G \, , \nonumber\\  
&\frac{d {\cal F}^G(y,\zeta,Q^2)}{d\ln(Q^2)}= \Big[\int^1_y \frac{dz}{\zeta} V^{gg}\left(\frac{y}{\zeta},  
\frac{z}{\zeta}\right)_+ +\nonumber\\  
&\int^y_0 \frac{dz}{\zeta} V^{gg}\left(\overline {\frac{y}{\zeta}},\overline {\frac{z}{\zeta}}\right)_+  
\pm \int^1_{\zeta} \frac{dz}{\zeta} V^{gg}\left(\overline{\frac{y}{\zeta}},\frac{z}{\zeta}\right)\Big]  
{\cal F}^G\nonumber\\  
&+\frac{1}{\left(1-\frac{\zeta}{2}\right)}\Big[\int^1_y dz  
V^{gq}\left(\frac{y}{\zeta},\frac{z}{\zeta}\right)\nonumber\\  
&- \int^y_0 dz V^{gq}\left(\overline {\frac{y}{\zeta}},\overline {\frac{z}{\zeta}}\right)  
\pm \int^1_{\zeta} dz V^{gq}\left(\overline{\frac{y}{\zeta}},\frac{z}{\zeta}\right)\Big] {\cal F}^S \, , \label{erbleq} 
\end{align} 
with generalized ERBL kernels \cite{bfm}, where the bar notation means 
for example, $\overline{z/\zeta} = 1 - z/\zeta$. The  
upper signs correspond to the unpolarized case and the lower signs to the polarized case.  

The numerical implementation of the $+$-distribution, again applied to the whole kernel, takes the following form in the ERBL region:  
\begin{align} 
&\int^1_y \frac{dz}{\zeta} V\left(\frac{y}{\zeta},\frac{z}{\zeta}\right)_+ {\cal F}(z,\zeta) = \nonumber\\  
&\int^1_y \frac{dz}{\zeta} V\left(\frac{y}{\zeta},\frac{z}{\zeta}\right)\left[ {\cal F}(z,\zeta)-{\cal F}(y,\zeta)\right]  + {\cal F}(y,\zeta) \times \nonumber\\  
& \left[ \int^{\zeta}_y \frac{dz}{\zeta} \left(V\left(\frac{y}{\zeta},\frac{z}{\zeta}\right) -  
V\left(\overline {\frac{z}{\zeta}},\overline {\frac{y}{\zeta}}\right)\right) 
+ \int^{1}_{\zeta} \frac{dz}{\zeta} V(\frac{y}{\zeta},\frac{z}{\zeta}) \right]
\, , \nonumber\\  
&\int^y_0 \frac{dz}{\zeta} V\left(\overline {\frac{y}{\zeta}},\overline {\frac{z}{\zeta}}\right)_+{\cal F}(z,\zeta) =  
\nonumber\\  
&\int^y_0 \frac{dz}{\zeta} V\left(\overline {\frac{y}{\zeta}},\overline {\frac{z}{\zeta}}\right)  
\left[{\cal F} (z,\zeta)-{\cal F}(y,\zeta)\right]\nonumber\\  
& + {\cal F}(y,\zeta)\int^y_0 \frac{dz}{\zeta} \left(V\left(\overline {\frac{y}{\zeta}},\overline {\frac{z}{\zeta}}\right)  
- V\left(\frac{z}{\zeta},\frac{y}{\zeta}\right)\right) \, ,  
\end{align} 
where the terms have been arranged in such away that all non-integrable divergences explicitly cancel in each term separately\cite{foot3}.  
 
\section{Results}  
  
\label{sec:res}  

We illustrate our polarized and unpolarized NLO GPDs in several ways, for two values of skewedness, $\zeta = 0.1, 0.0001$, 
which are representative of the kinematic ranges accessible in HERMES and HERA kinematics, respectively. 
In particular, we quantify the effect of skewed evolution at NLO in the DGLAP region by comparing to conventional NLO DGLAP evolution of the same inputs, we present the relative influence of NLO skewed kernels by plotting the ratio of  
NLO to LO skewed evolution of common inputs and we also present the NLO GPDs themselves.  
In addition, we quantify the significance of the D-term and investigate the effect of changing the profile 
functions of eq.(\ref{ddinp}).
 
\begin{figure} 
\centering 
\mbox{\epsfig{file=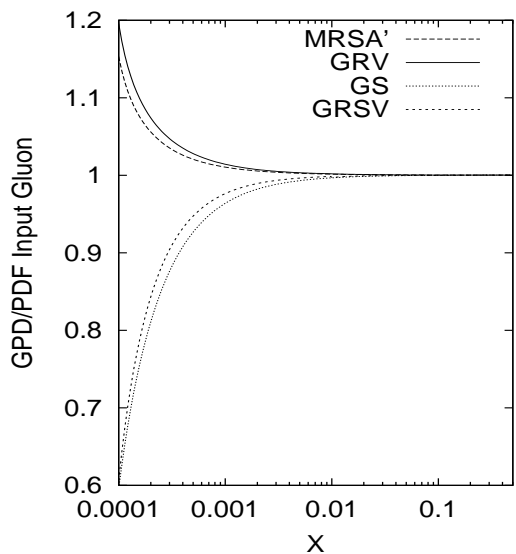,width=7.5cm,height=7.5cm}} 
\caption{The ratio of the unpolarized and polarized input GPDs to conventional PDF  
inputs in the DGLAP region, for $\zeta = 0.0001$.  
The forward limit GPD/PDF $\to 1$, for $X \gg \zeta$ is observed in each case.  
The unpolarized GPDs are enhanced and the polarized GPDs suppressed close to $X = \zeta=0.0001$.} 
\label{fig:ratfwd} 
\end{figure} 
  
The skewed evolution of eqs.(\ref{dglapeq},\ref{erbleq}) of the input distributions defined in Section.(\ref{sec:inp}), for a particular 
$\zeta$, is implemented by numerical integration on a fine grid in $X^{'}$ and $Q^{'}$, using 
the CTEQ \cite{cteqcode} evolution package, for initialization of the input and implementation of the $Q^2$ evolution. The convolution piece was completely rewritten. All code is written in Fortran77 and will be made available via the internet \cite{website}. 

\subsection{Comparison of skewed to conventional DGLAP evolution at NLO} 

\begin{figure} 
\centering 
\mbox{\epsfig{file=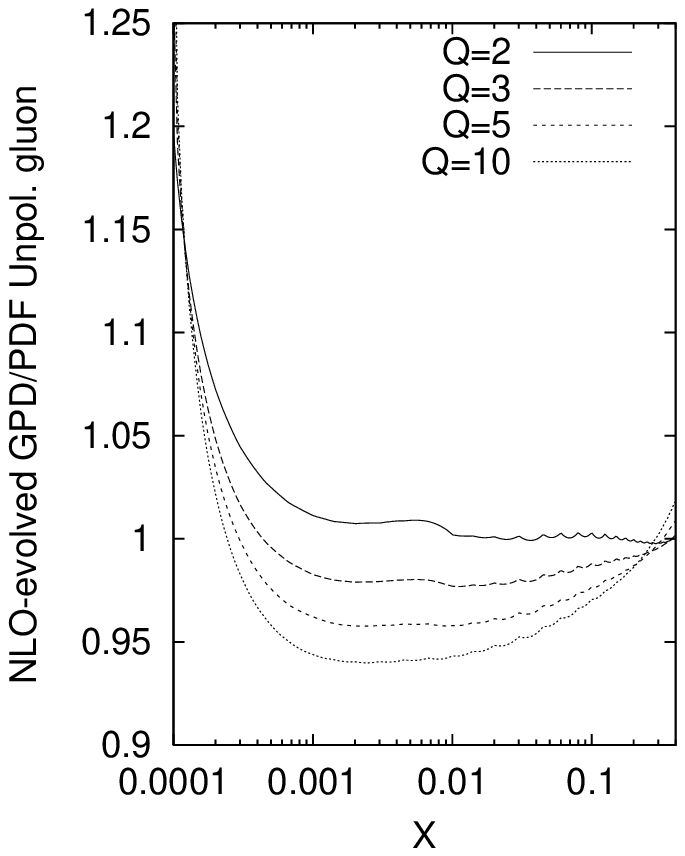,width=7.5cm,height=6.5cm}}
\end{figure} 
\begin{figure} 
\centering 
\mbox{\epsfig{file=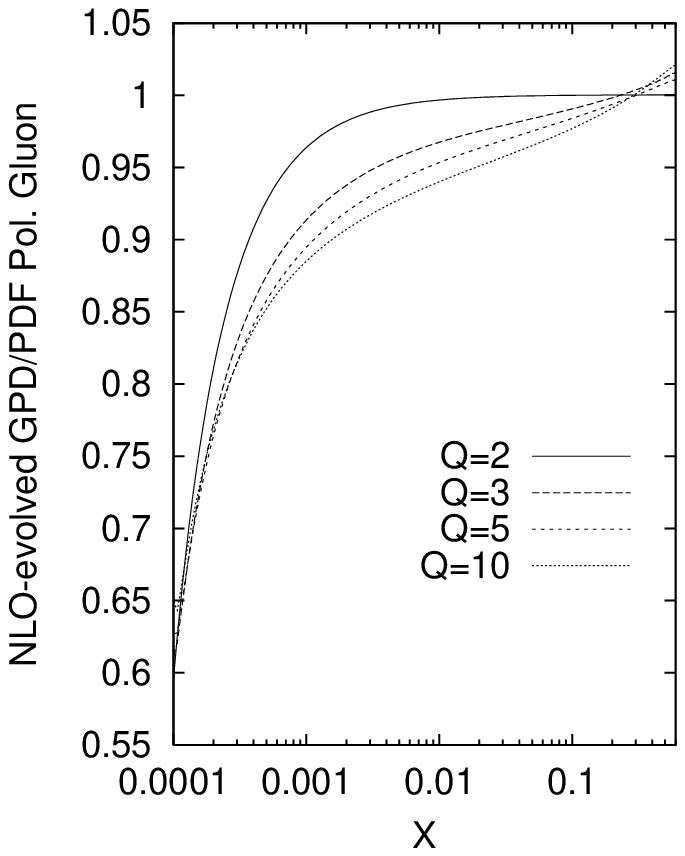,width=7.5cm,height=6.5cm}} 
\caption{Ratio of NLO-evolved GPD gluon over conventional PDF gluon at the input 
scale and several values of $Q$, for $\zeta=0.0001$. The upper plot is for the unpolarized 
case (GRV98) and the lower one for the polarized case (GS(A)).} 
\label{fig:ratunp} 
\end{figure} 

In fig.(\ref{fig:ratfwd}) we plot the ratio of our input gluon GPDs to the conventional input gluon PDFs on which they are based,  
in the DGLAP region ($X> \zeta = 0.0001$).  The forward limit ${\cal F} (X \gg \zeta,\zeta,Q_0) / f (X,Q_0) \to 1$,  
built into our input model of eqs.(\ref{fqinp}, \ref{ddinp}, \ref{pdfinp}) by design, is illustrated for each input PDF.  
In the region $X \approx \zeta$, the unpolarized distributions MRSA$^{'}$ and GRV98 exhibit the familiar enhancement (of about $ 15-20 \%$) required by positivity constraints,  
${\cal F} (X \approx \zeta,\zeta,Q_0) / f (X,Q_0) > 1$. In contrast, the polarized input model  
shows a suppression (of about $35-40 \%$) in this region. 
We choose to plot results for the gluon since this is most relevant to small-$x$ exclusive 
diffractive processes at HERA, for which it has been argued that the inclusion of skewedness (at least at LO in the unpolarized case) leads to an enhancement (see e.g. \cite{vgaf,kgbm,fms}). 

In \cite{vgaf} it was demonstrated that LO skewed evolution in the unpolarized case yields  
a ratio consistently above unity with a strong enhancement near $X=\zeta$ and a ratio tending to unity from above at large $X \gg \zeta$. This is due to the probabilistic interpretation of the  
evolution at LO \cite{tery}. At NLO, this interpretation is lost in the unpolarized case \cite{foot4}  
and thus the LO statement does not necessarily apply, since perturbative quantities become scheme dependent in NLO.  

\begin{figure}  
\centering  
\mbox{\epsfig{file=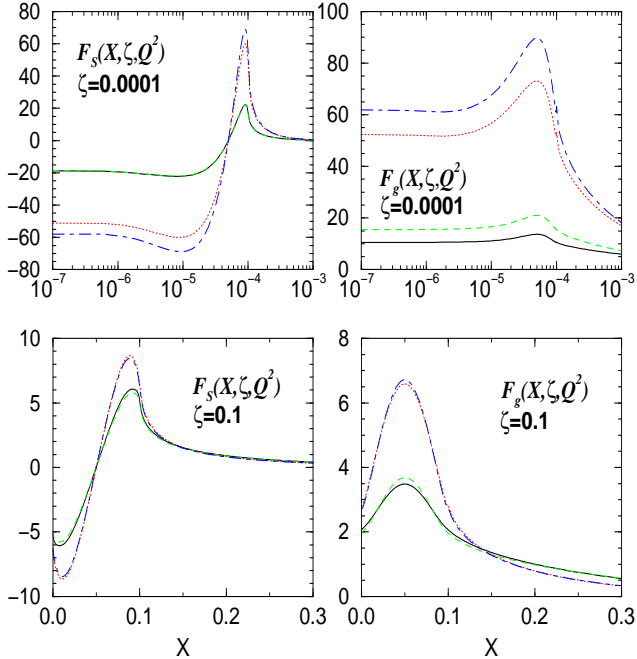,width=8.5cm,height=9cm}}  
\caption{Unpolarized NLO input and evolved singlet quark and gluon GPDs at small and large skewedness. 
The solid curves are the input MRSA$^{'}$ GPDs at $Q_0 = 2~\mbox{GeV}$, the dotted ones show them evolved to $Q =10~\mbox{GeV}$.  
The dashed curves are input GRV98 GPDs at $Q_0 = 2~\mbox{GeV}$ and the dashed-dotted ones show them evolved to $Q = 10~\mbox{GeV}$. 
The quark singlet is scaled by a factor of $10^{-4}$ at $\zeta = 0.0001$ and by $10^{-2}$ at $\zeta = 0.1$. 
For $\zeta=0.1$ the symmetry of the gluon GPD and the anti-symmetry of the singlet quark GPD are apparent about the point $X=\zeta/2 = 0.05$.} 
\label{fig:evolveall} 
\end{figure}  
 
In the upper plot of fig.(\ref{fig:ratunp}) we show the same ratio for the unpolarized  
evolved gluon distribution of GRV98 \cite{foot5}. It is interesting and somewhat surprising to observe that this ratio drops progressively below unity even for $X \gg \zeta$ as $Q^2$ increases, and is as much as $6\%$ below at $X=0.001$ for $Q=10$~GeV. However, on inspecting the ratio of the NLO $gg$ and $gq$ kernels for the GPDs to the NLO $gg$ and $gq$ kernels of the forward PDFs, one observes that even for $X \gg \zeta$ they can be very different from one another over a broad range of $X$, but will eventually settle at unity. An enhancement due to skewed evolution is still seen at NLO , but only very close to $X = \zeta$ (for both the polarized and unpolarized cases). 
 
In the lower plot of fig.(\ref{fig:ratunp}) we illustrate the
polarized case using the GS(A) polarized gluon. Again we see the ratio drop progressively below unity for $X> \zeta$. The size of the effect is similar, although this is difficult to see directly since the ratio of GPD to PDF is already smaller than unity at the input scale (for $X \approx \zeta$) \cite{foot6}.  
 
\subsection{NLO GPDs and a comparison with LO} 

\begin{figure}  
\centering  
\mbox{\epsfig{file=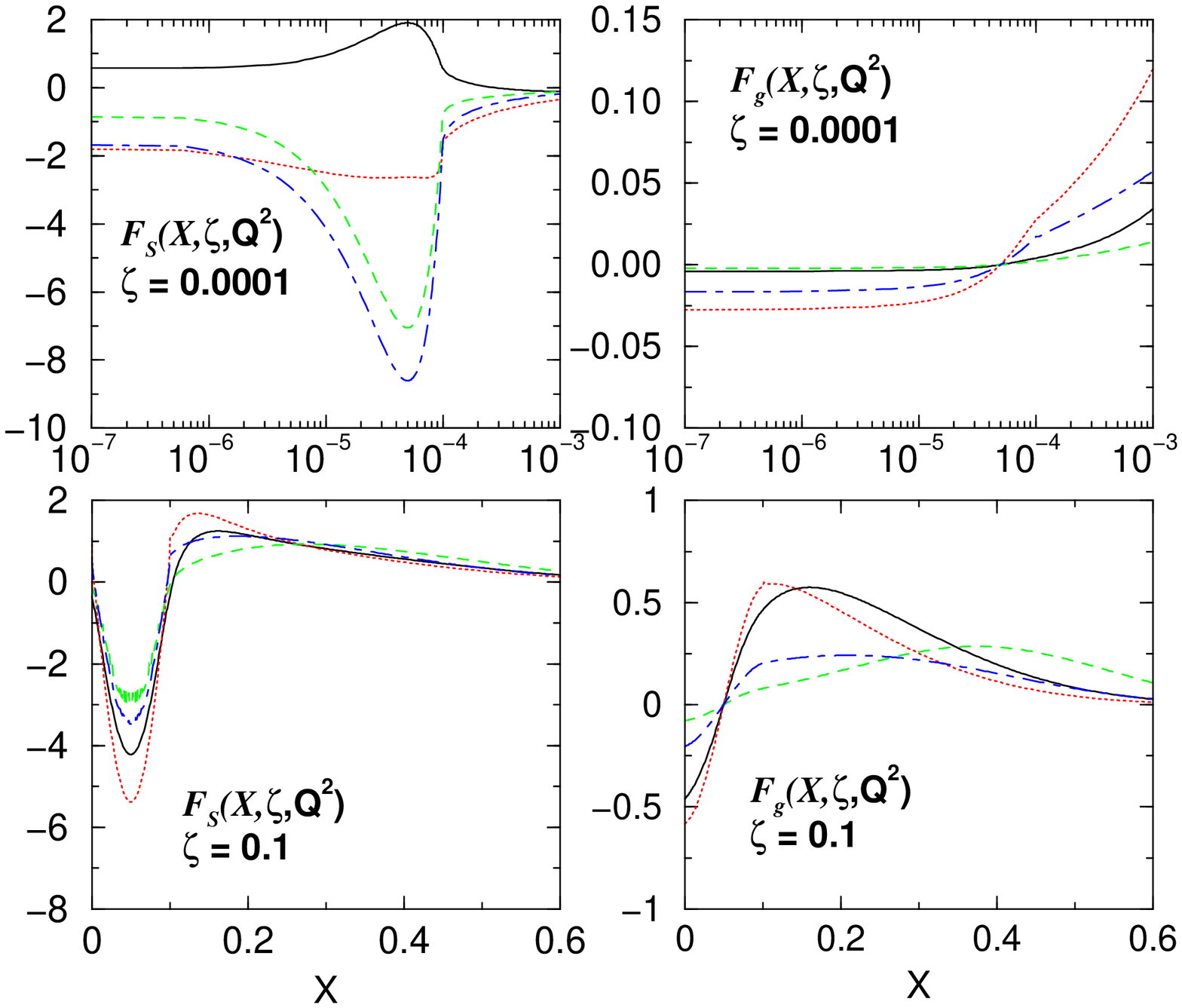,width=8.5cm,height=8.75cm}}  
\caption{Polarized NLO input and evolved singlet quark and gluon GPDs at small and large skewedness. 
The solid curves are the input GS(A) GPDs at $Q_0 = 2~\mbox{GeV}$, the dotted ones show them evolved to $Q =10~\mbox{GeV}$.  
The dashed curves are input GRSV00 GPDs at $Q_0 = 2~\mbox{GeV}$ and the dashed-dotted ones show them evolved to $Q = 10~\mbox{GeV}$. 
The quark singlet is scaled by a factor of $10^{-2}$ at $\zeta = 0.0001$. 
For $\zeta=0.1$ the symmetry of the polarized singlet quark GPD and the anti-symmetry of the polarized gluon GPD are apparent about the point $X=\zeta/2 = 0.05$.} 
\label{fig:evolveallpol} 
\end{figure} 

\begin{figure}  
\centering  
\mbox{\epsfig{file=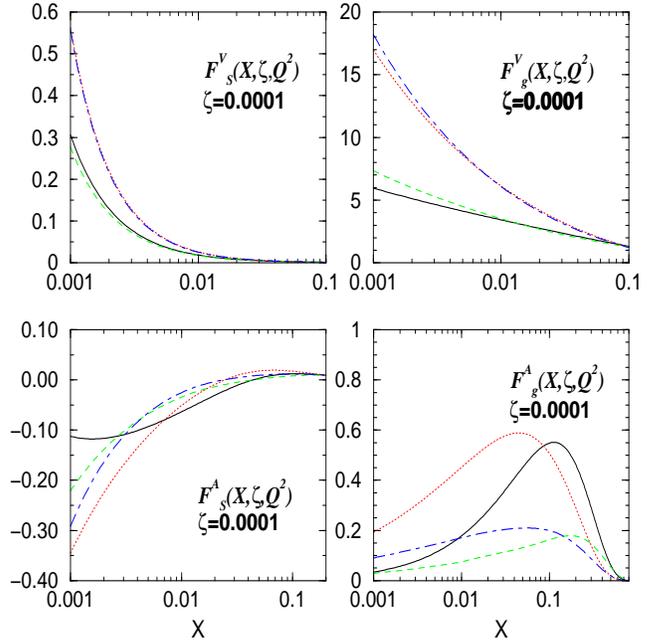,width=8.5cm,height=8.75cm}}  
\caption{Polarised (``A'') and unpolarised (``V'') singlet quark and gluon GPDs for 
$\zeta=0.0001$ in the $X > 10 \zeta$ region. The input parton sets for each curve are the same as those given in Figs. \ref{fig:evolveall} and \ref{fig:evolveallpol}. The quark singlet is scaled by a factor of $10^{-4}$ in the unpolarised case 
and by $10^{-2}$ in the polarised case.} 
\label{fig:evolveall1} 
\end{figure}

In figs.(\ref{fig:evolveall},\ref{fig:evolveallpol}, \ref{fig:evolveall1}) we present the input and evolved unpolarized  
and polarized GPDs, respectively, for the singlet quark and gluon, and are encouraged to note that they are smooth, 
 continuous functions of $X$. 
The lower plots in figs.(\ref{fig:evolveall},\ref{fig:evolveallpol}), with $\zeta=0.1$, use a linear scale on the $X$ axis, 
and hence illustrate very explicitly the symmetries in the ERBL region about the point $X= \zeta/2 = 0.05$ which are 
manifestly preserved under evolution.  
These symmetries also hold for the upper curves (for $\zeta=0.0001$) but are less visible on a logarithmic scale.  
These features indicate that our code is indeed correct and properly performs the NLO evolution in both the 
ERBL and DGLAP region. 
From fig.(\ref{fig:evolveall}) and the upper plots of fig.(\ref{fig:evolveall1}) we note that both unpolarised 
input sets (GRV98 and MRSA') produce similar results for the input and evolved GPDs.

From fig.(\ref{fig:evolveallpol}) and the lower left plot of fig.(\ref{fig:evolveall1}) we note with 
particular interest the result for the NLO evolution of the polarized quark singlet at small $\zeta$. 
The GS(A) set assumes a rather different sea content to GRSV00, in particular the former has a  
positive quark singlet at the input scale and the latter a negative one. 
The NLO evolution keeps the GRSV00 distribution unchanged in shape, 
only altering its size, whereas the 
GS(A) distribution is driven towards the same shape as well as overall  
sign as the GRSV00 distribution. This seems to indicate that the NLO 
generalised evolution favours the initial shape of the GRSV00 distribution and  
thus a negative polarized quark singlet at small $\zeta$. 
At larger $\zeta$, where the sea is of little numerical importance, the 
initial shapes are very similar and so one does not observe the same type of  
effect for GS(A). 

Although similar in shape (except for the polarized quark singlet at the input), the two sets of polarised distributions  
differ considerably in size. The general evolution effects which push partons to smaller $X$ with increasing $Q^2$,  
are naturally the same for both sets of inputs. 
 
\begin{figure}  
\centering  
\mbox{\epsfig{file=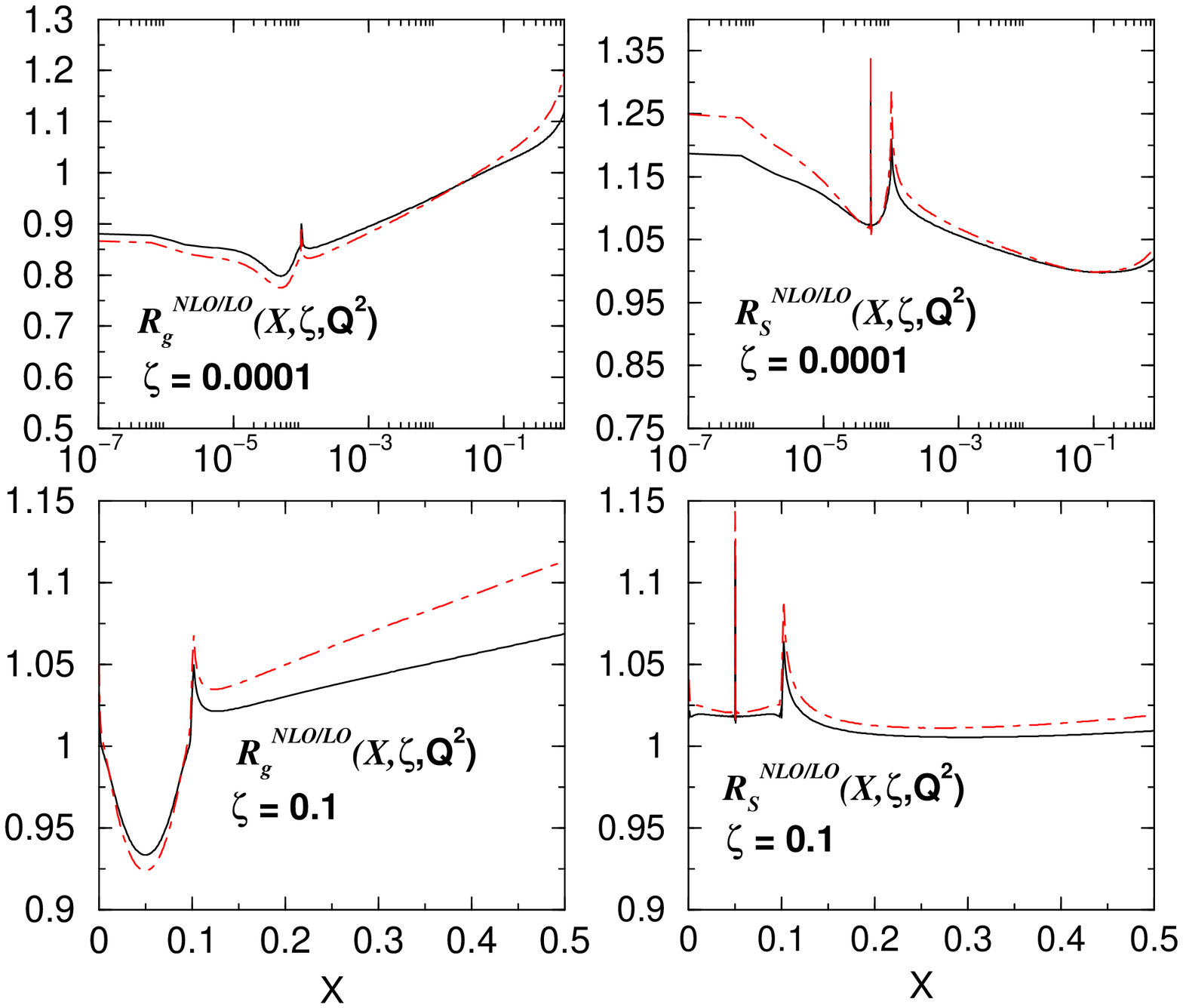,width=7.5cm,height=7.5cm}}  
\mbox{\epsfig{file=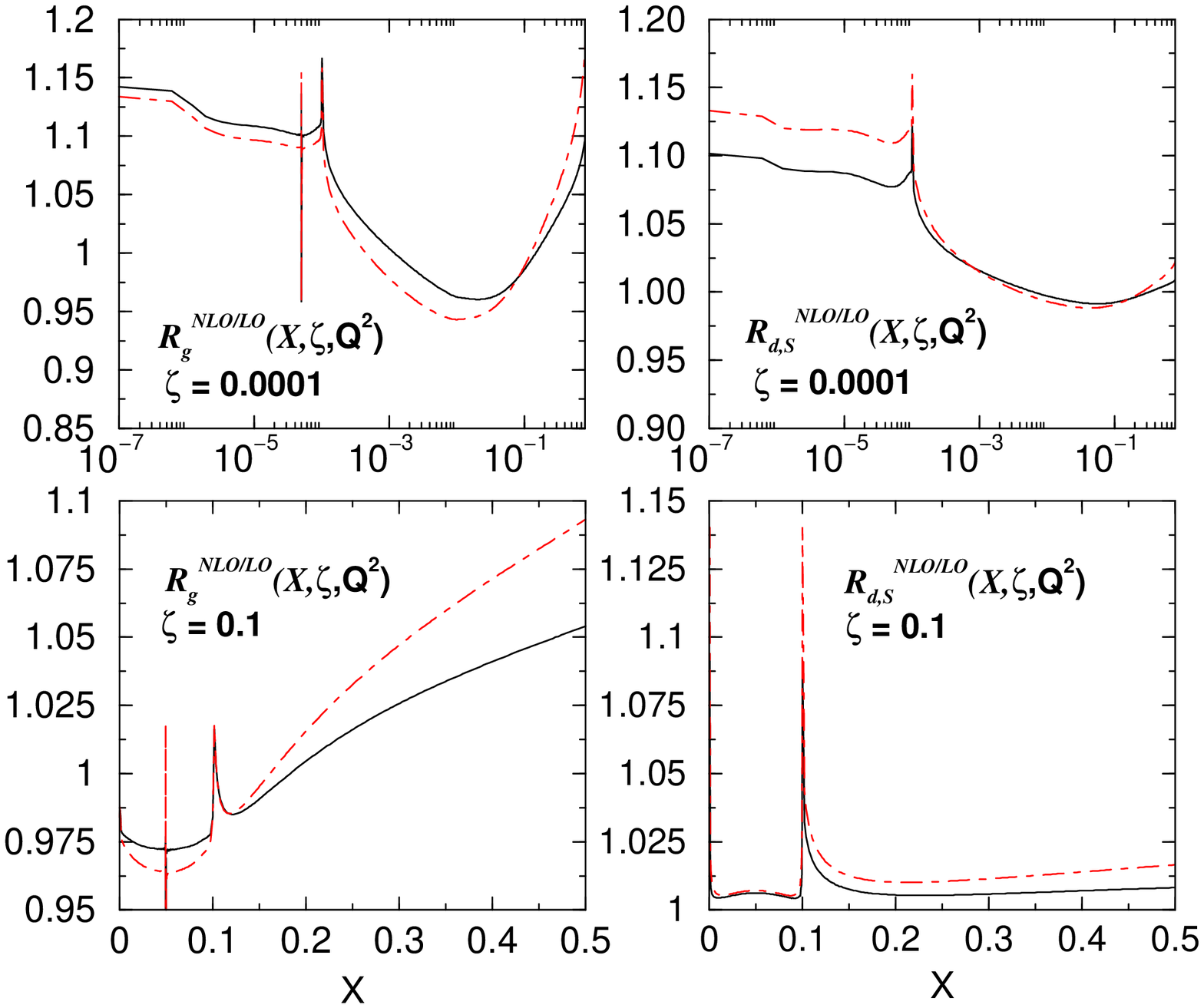,width=7.5cm,height=7.5cm}} 
\caption{The ratio of NLO to LO evolved GPDs for the unpolarized
(upper plot, MRSA$^{'}$) and polarized (lower plot, 
GRSV) cases. The solid and dashed-dotted curves shows the ratio at
$Q=5~\mbox{GeV}$ and $Q=10~\mbox{GeV}$ , 
respectively.} 
\label{fig:nloloev} 
\end{figure}

In order to quantify the relative effects of including NLO skewed kernels in detail, we also evolved the same input GPDs using LO skewed kernels. In the upper and lower plots of fig.(\ref{fig:nloloev}) we present the ratio of the evolved GPDs at NLO to those at LO, for the  
unpolarized and polarized cases, respectively. Overall the effect of including NLO corrections is at most a $25\%$ effect, within the kinematic ranges illustrated. The regions furthest from unity  
in these ratio plots illustrate the ranges in $X$, relative to $\zeta$, where the greatest effects are felt. 
The symmetry of the individual distributions about $\zeta/2$ continue to be  
reflected in these ratio plots, which further indicates that the NLO evolution slightly alters 
the shape of the distributions relative to LO, but in such a way as to preserve the symmetries.  

The spikes in the antisymmetric unpolarized singlet quark and polarized gluon ratios  
result from a slight loss of numerical accuracy in a few bins around the 
point $\zeta/2$ (where the distributions have to pass through zero). The effect is magnified in this ratio. 
However, since the distributions vary by as much as twelve orders of magnitude (in the unpolarized quark singlet case),  
and the effect occurs close to a zero, this is not a cause for concern. 
 
The cusps observed at $X=\zeta$ are due to the fact that  
the NLO over LO ratios approach the same value from the ERBL and DGLAP region very rapidly  
around the $X=\zeta$ region.  
This change in shape in NLO over LO, which could be seen more clearly if 
the region around the $X=\zeta$ were to be enlarged, is easily explained.  
In the LO kernels, rational functions of the type  
$1/(X-\zeta)$ appear, which however combine to form a function which is 
finite at the point $X=\zeta$. In NLO, functions 
of the type $\ln^n(X-\zeta)/(X-\zeta)^{n_1}$ with $n,n_1=0,1,2$ appear,  
where again the rational functions cancel, but the logarithms do not.  
This leads to an integrable logarithmic singularity at NLO of a similar type as encountered in the NLO  
forward evolution, at the lower bound of the convolution integral in the RGE, $y=x_{bj}$, and leads to a strong 
enhancement of the NLO distributions in the region around $X=\zeta$ as seen in the figures.  
The ratio of NLO to LO in the unpolarized case for GRV98 are very 
similar, within a few percent, to the plotted MRSA$^{'}$. In the 
polarized case the contrast is larger, until very large $Q^2$, due to the 
markedly different input functions used for GRSV and GS(A)\cite{nlolocomment}.  
 
\subsection{Influence of the D-term} 
 
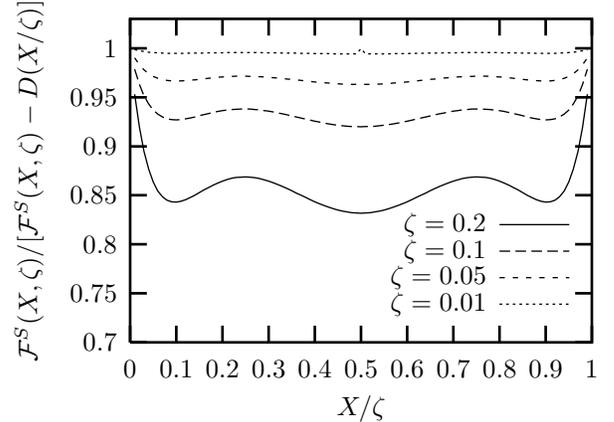
\begin{figure}  
\begingroup%
  \makeatletter%
  \newcommand{\GNUPLOTspecial}{%
    \@sanitize\catcode`\%=14\relax\special}%
  \setlength{\unitlength}{0.1bp}%
{\GNUPLOTspecial{!
/gnudict 256 dict def
gnudict begin
/Color false def
/Solid false def
/gnulinewidth 5.000 def
/userlinewidth gnulinewidth def
/vshift -33 def
/dl {10 mul} def
/hpt_ 31.5 def
/vpt_ 31.5 def
/hpt hpt_ def
/vpt vpt_ def
/M {moveto} bind def
/L {lineto} bind def
/R {rmoveto} bind def
/V {rlineto} bind def
/vpt2 vpt 2 mul def
/hpt2 hpt 2 mul def
/Lshow { currentpoint stroke M
  0 vshift R show } def
/Rshow { currentpoint stroke M
  dup stringwidth pop neg vshift R show } def
/Cshow { currentpoint stroke M
  dup stringwidth pop -2 div vshift R show } def
/UP { dup vpt_ mul /vpt exch def hpt_ mul /hpt exch def
  /hpt2 hpt 2 mul def /vpt2 vpt 2 mul def } def
/DL { Color {setrgbcolor Solid {pop []} if 0 setdash }
 {pop pop pop Solid {pop []} if 0 setdash} ifelse } def
/BL { stroke gnulinewidth 2 mul setlinewidth } def
/AL { stroke gnulinewidth 2 div setlinewidth } def
/UL { gnulinewidth mul /userlinewidth exch def } def
/PL { stroke userlinewidth setlinewidth } def
/LTb { BL [] 0 0 0 DL } def
/LTa { AL [1 dl 2 dl] 0 setdash 0 0 0 setrgbcolor } def
/LT0 { PL [] 1 0 0 DL } def
/LT1 { PL [4 dl 2 dl] 0 1 0 DL } def
/LT2 { PL [2 dl 3 dl] 0 0 1 DL } def
/LT3 { PL [1 dl 1.5 dl] 1 0 1 DL } def
/LT4 { PL [5 dl 2 dl 1 dl 2 dl] 0 1 1 DL } def
/LT5 { PL [4 dl 3 dl 1 dl 3 dl] 1 1 0 DL } def
/LT6 { PL [2 dl 2 dl 2 dl 4 dl] 0 0 0 DL } def
/LT7 { PL [2 dl 2 dl 2 dl 2 dl 2 dl 4 dl] 1 0.3 0 DL } def
/LT8 { PL [2 dl 2 dl 2 dl 2 dl 2 dl 2 dl 2 dl 4 dl] 0.5 0.5 0.5 DL } def
/Pnt { stroke [] 0 setdash
   gsave 1 setlinecap M 0 0 V stroke grestore } def
/Dia { stroke [] 0 setdash 2 copy vpt add M
  hpt neg vpt neg V hpt vpt neg V
  hpt vpt V hpt neg vpt V closepath stroke
  Pnt } def
/Pls { stroke [] 0 setdash vpt sub M 0 vpt2 V
  currentpoint stroke M
  hpt neg vpt neg R hpt2 0 V stroke
  } def
/Box { stroke [] 0 setdash 2 copy exch hpt sub exch vpt add M
  0 vpt2 neg V hpt2 0 V 0 vpt2 V
  hpt2 neg 0 V closepath stroke
  Pnt } def
/Crs { stroke [] 0 setdash exch hpt sub exch vpt add M
  hpt2 vpt2 neg V currentpoint stroke M
  hpt2 neg 0 R hpt2 vpt2 V stroke } def
/TriU { stroke [] 0 setdash 2 copy vpt 1.12 mul add M
  hpt neg vpt -1.62 mul V
  hpt 2 mul 0 V
  hpt neg vpt 1.62 mul V closepath stroke
  Pnt  } def
/Star { 2 copy Pls Crs } def
/BoxF { stroke [] 0 setdash exch hpt sub exch vpt add M
  0 vpt2 neg V  hpt2 0 V  0 vpt2 V
  hpt2 neg 0 V  closepath fill } def
/TriUF { stroke [] 0 setdash vpt 1.12 mul add M
  hpt neg vpt -1.62 mul V
  hpt 2 mul 0 V
  hpt neg vpt 1.62 mul V closepath fill } def
/TriD { stroke [] 0 setdash 2 copy vpt 1.12 mul sub M
  hpt neg vpt 1.62 mul V
  hpt 2 mul 0 V
  hpt neg vpt -1.62 mul V closepath stroke
  Pnt  } def
/TriDF { stroke [] 0 setdash vpt 1.12 mul sub M
  hpt neg vpt 1.62 mul V
  hpt 2 mul 0 V
  hpt neg vpt -1.62 mul V closepath fill} def
/DiaF { stroke [] 0 setdash vpt add M
  hpt neg vpt neg V hpt vpt neg V
  hpt vpt V hpt neg vpt V closepath fill } def
/Pent { stroke [] 0 setdash 2 copy gsave
  translate 0 hpt M 4 {72 rotate 0 hpt L} repeat
  closepath stroke grestore Pnt } def
/PentF { stroke [] 0 setdash gsave
  translate 0 hpt M 4 {72 rotate 0 hpt L} repeat
  closepath fill grestore } def
/Circle { stroke [] 0 setdash 2 copy
  hpt 0 360 arc stroke Pnt } def
/CircleF { stroke [] 0 setdash hpt 0 360 arc fill } def
/C0 { BL [] 0 setdash 2 copy moveto vpt 90 450  arc } bind def
/C1 { BL [] 0 setdash 2 copy        moveto
       2 copy  vpt 0 90 arc closepath fill
               vpt 0 360 arc closepath } bind def
/C2 { BL [] 0 setdash 2 copy moveto
       2 copy  vpt 90 180 arc closepath fill
               vpt 0 360 arc closepath } bind def
/C3 { BL [] 0 setdash 2 copy moveto
       2 copy  vpt 0 180 arc closepath fill
               vpt 0 360 arc closepath } bind def
/C4 { BL [] 0 setdash 2 copy moveto
       2 copy  vpt 180 270 arc closepath fill
               vpt 0 360 arc closepath } bind def
/C5 { BL [] 0 setdash 2 copy moveto
       2 copy  vpt 0 90 arc
       2 copy moveto
       2 copy  vpt 180 270 arc closepath fill
               vpt 0 360 arc } bind def
/C6 { BL [] 0 setdash 2 copy moveto
      2 copy  vpt 90 270 arc closepath fill
              vpt 0 360 arc closepath } bind def
/C7 { BL [] 0 setdash 2 copy moveto
      2 copy  vpt 0 270 arc closepath fill
              vpt 0 360 arc closepath } bind def
/C8 { BL [] 0 setdash 2 copy moveto
      2 copy vpt 270 360 arc closepath fill
              vpt 0 360 arc closepath } bind def
/C9 { BL [] 0 setdash 2 copy moveto
      2 copy  vpt 270 450 arc closepath fill
              vpt 0 360 arc closepath } bind def
/C10 { BL [] 0 setdash 2 copy 2 copy moveto vpt 270 360 arc closepath fill
       2 copy moveto
       2 copy vpt 90 180 arc closepath fill
               vpt 0 360 arc closepath } bind def
/C11 { BL [] 0 setdash 2 copy moveto
       2 copy  vpt 0 180 arc closepath fill
       2 copy moveto
       2 copy  vpt 270 360 arc closepath fill
               vpt 0 360 arc closepath } bind def
/C12 { BL [] 0 setdash 2 copy moveto
       2 copy  vpt 180 360 arc closepath fill
               vpt 0 360 arc closepath } bind def
/C13 { BL [] 0 setdash  2 copy moveto
       2 copy  vpt 0 90 arc closepath fill
       2 copy moveto
       2 copy  vpt 180 360 arc closepath fill
               vpt 0 360 arc closepath } bind def
/C14 { BL [] 0 setdash 2 copy moveto
       2 copy  vpt 90 360 arc closepath fill
               vpt 0 360 arc } bind def
/C15 { BL [] 0 setdash 2 copy vpt 0 360 arc closepath fill
               vpt 0 360 arc closepath } bind def
/Rec   { newpath 4 2 roll moveto 1 index 0 rlineto 0 exch rlineto
       neg 0 rlineto closepath } bind def
/Square { dup Rec } bind def
/Bsquare { vpt sub exch vpt sub exch vpt2 Square } bind def
/S0 { BL [] 0 setdash 2 copy moveto 0 vpt rlineto BL Bsquare } bind def
/S1 { BL [] 0 setdash 2 copy vpt Square fill Bsquare } bind def
/S2 { BL [] 0 setdash 2 copy exch vpt sub exch vpt Square fill Bsquare } bind def
/S3 { BL [] 0 setdash 2 copy exch vpt sub exch vpt2 vpt Rec fill Bsquare } bind def
/S4 { BL [] 0 setdash 2 copy exch vpt sub exch vpt sub vpt Square fill Bsquare } bind def
/S5 { BL [] 0 setdash 2 copy 2 copy vpt Square fill
       exch vpt sub exch vpt sub vpt Square fill Bsquare } bind def
/S6 { BL [] 0 setdash 2 copy exch vpt sub exch vpt sub vpt vpt2 Rec fill Bsquare } bind def
/S7 { BL [] 0 setdash 2 copy exch vpt sub exch vpt sub vpt vpt2 Rec fill
       2 copy vpt Square fill
       Bsquare } bind def
/S8 { BL [] 0 setdash 2 copy vpt sub vpt Square fill Bsquare } bind def
/S9 { BL [] 0 setdash 2 copy vpt sub vpt vpt2 Rec fill Bsquare } bind def
/S10 { BL [] 0 setdash 2 copy vpt sub vpt Square fill 2 copy exch vpt sub exch vpt Square fill
       Bsquare } bind def
/S11 { BL [] 0 setdash 2 copy vpt sub vpt Square fill 2 copy exch vpt sub exch vpt2 vpt Rec fill
       Bsquare } bind def
/S12 { BL [] 0 setdash 2 copy exch vpt sub exch vpt sub vpt2 vpt Rec fill Bsquare } bind def
/S13 { BL [] 0 setdash 2 copy exch vpt sub exch vpt sub vpt2 vpt Rec fill
       2 copy vpt Square fill Bsquare } bind def
/S14 { BL [] 0 setdash 2 copy exch vpt sub exch vpt sub vpt2 vpt Rec fill
       2 copy exch vpt sub exch vpt Square fill Bsquare } bind def
/S15 { BL [] 0 setdash 2 copy Bsquare fill Bsquare } bind def
/D0 { gsave translate 45 rotate 0 0 S0 stroke grestore } bind def
/D1 { gsave translate 45 rotate 0 0 S1 stroke grestore } bind def
/D2 { gsave translate 45 rotate 0 0 S2 stroke grestore } bind def
/D3 { gsave translate 45 rotate 0 0 S3 stroke grestore } bind def
/D4 { gsave translate 45 rotate 0 0 S4 stroke grestore } bind def
/D5 { gsave translate 45 rotate 0 0 S5 stroke grestore } bind def
/D6 { gsave translate 45 rotate 0 0 S6 stroke grestore } bind def
/D7 { gsave translate 45 rotate 0 0 S7 stroke grestore } bind def
/D8 { gsave translate 45 rotate 0 0 S8 stroke grestore } bind def
/D9 { gsave translate 45 rotate 0 0 S9 stroke grestore } bind def
/D10 { gsave translate 45 rotate 0 0 S10 stroke grestore } bind def
/D11 { gsave translate 45 rotate 0 0 S11 stroke grestore } bind def
/D12 { gsave translate 45 rotate 0 0 S12 stroke grestore } bind def
/D13 { gsave translate 45 rotate 0 0 S13 stroke grestore } bind def
/D14 { gsave translate 45 rotate 0 0 S14 stroke grestore } bind def
/D15 { gsave translate 45 rotate 0 0 S15 stroke grestore } bind def
/DiaE { stroke [] 0 setdash vpt add M
  hpt neg vpt neg V hpt vpt neg V
  hpt vpt V hpt neg vpt V closepath stroke } def
/BoxE { stroke [] 0 setdash exch hpt sub exch vpt add M
  0 vpt2 neg V hpt2 0 V 0 vpt2 V
  hpt2 neg 0 V closepath stroke } def
/TriUE { stroke [] 0 setdash vpt 1.12 mul add M
  hpt neg vpt -1.62 mul V
  hpt 2 mul 0 V
  hpt neg vpt 1.62 mul V closepath stroke } def
/TriDE { stroke [] 0 setdash vpt 1.12 mul sub M
  hpt neg vpt 1.62 mul V
  hpt 2 mul 0 V
  hpt neg vpt -1.62 mul V closepath stroke } def
/PentE { stroke [] 0 setdash gsave
  translate 0 hpt M 4 {72 rotate 0 hpt L} repeat
  closepath stroke grestore } def
/CircE { stroke [] 0 setdash 
  hpt 0 360 arc stroke } def
/Opaque { gsave closepath 1 setgray fill grestore 0 setgray closepath } def
/DiaW { stroke [] 0 setdash vpt add M
  hpt neg vpt neg V hpt vpt neg V
  hpt vpt V hpt neg vpt V Opaque stroke } def
/BoxW { stroke [] 0 setdash exch hpt sub exch vpt add M
  0 vpt2 neg V hpt2 0 V 0 vpt2 V
  hpt2 neg 0 V Opaque stroke } def
/TriUW { stroke [] 0 setdash vpt 1.12 mul add M
  hpt neg vpt -1.62 mul V
  hpt 2 mul 0 V
  hpt neg vpt 1.62 mul V Opaque stroke } def
/TriDW { stroke [] 0 setdash vpt 1.12 mul sub M
  hpt neg vpt 1.62 mul V
  hpt 2 mul 0 V
  hpt neg vpt -1.62 mul V Opaque stroke } def
/PentW { stroke [] 0 setdash gsave
  translate 0 hpt M 4 {72 rotate 0 hpt L} repeat
  Opaque stroke grestore } def
/CircW { stroke [] 0 setdash 
  hpt 0 360 arc Opaque stroke } def
/BoxFill { gsave Rec 1 setgray fill grestore } def
end
}}%
\begin{picture}(2340,1620)(0,0)%
{\GNUPLOTspecial{"
gnudict begin
gsave
0 0 translate
0.100 0.100 scale
0 setgray
newpath
1.000 UL
LTb
450 300 M
63 0 V
1677 0 R
-63 0 V
450 485 M
63 0 V
1677 0 R
-63 0 V
450 670 M
63 0 V
1677 0 R
-63 0 V
450 855 M
63 0 V
1677 0 R
-63 0 V
450 1039 M
63 0 V
1677 0 R
-63 0 V
450 1224 M
63 0 V
1677 0 R
-63 0 V
450 1409 M
63 0 V
1677 0 R
-63 0 V
450 300 M
0 63 V
0 1157 R
0 -63 V
624 300 M
0 63 V
0 1157 R
0 -63 V
798 300 M
0 63 V
0 1157 R
0 -63 V
972 300 M
0 63 V
0 1157 R
0 -63 V
1146 300 M
0 63 V
0 1157 R
0 -63 V
1320 300 M
0 63 V
0 1157 R
0 -63 V
1494 300 M
0 63 V
0 1157 R
0 -63 V
1668 300 M
0 63 V
0 1157 R
0 -63 V
1842 300 M
0 63 V
0 1157 R
0 -63 V
2016 300 M
0 63 V
0 1157 R
0 -63 V
2190 300 M
0 63 V
0 1157 R
0 -63 V
1.000 UL
LTb
450 300 M
1740 0 V
0 1220 V
-1740 0 V
450 300 L
1.000 UL
LT0
1842 744 M
263 0 V
467 1236 M
18 -124 V
17 -91 V
18 -68 V
17 -48 V
17 -34 V
18 -22 V
17 -13 V
18 -6 V
17 -1 V
17 3 V
18 6 V
17 8 V
18 9 V
17 10 V
17 10 V
18 9 V
17 9 V
18 8 V
17 7 V
17 6 V
18 5 V
17 3 V
18 1 V
17 1 V
17 -1 V
18 -2 V
17 -3 V
18 -5 V
17 -5 V
17 -6 V
18 -7 V
17 -7 V
18 -8 V
17 -8 V
17 -9 V
18 -8 V
17 -8 V
18 -9 V
17 -8 V
17 -7 V
18 -7 V
17 -6 V
18 -6 V
17 -5 V
17 -4 V
18 -3 V
17 -3 V
18 -1 V
17 -1 V
17 1 V
18 1 V
17 3 V
18 3 V
17 4 V
17 5 V
18 6 V
17 6 V
18 7 V
17 7 V
17 8 V
18 9 V
17 8 V
18 8 V
17 9 V
17 8 V
18 8 V
17 7 V
18 7 V
17 6 V
17 5 V
18 5 V
17 3 V
18 2 V
17 1 V
17 -1 V
18 -1 V
17 -3 V
18 -5 V
17 -6 V
17 -7 V
18 -8 V
17 -9 V
18 -9 V
17 -10 V
17 -10 V
18 -9 V
17 -8 V
18 -6 V
17 -3 V
17 1 V
18 6 V
17 13 V
18 22 V
17 34 V
17 48 V
18 68 V
17 91 V
18 124 V
1.000 UL
LT1
1842 644 M
263 0 V
467 1330 M
18 -57 V
17 -43 V
18 -31 V
17 -23 V
17 -15 V
18 -11 V
17 -7 V
18 -3 V
17 -1 V
17 1 V
18 3 V
17 3 V
18 4 V
17 5 V
17 4 V
18 4 V
17 4 V
18 4 V
17 3 V
17 2 V
18 2 V
17 1 V
18 1 V
17 0 V
17 0 V
18 -2 V
17 -1 V
18 -3 V
17 -2 V
17 -3 V
18 -4 V
17 -3 V
18 -4 V
17 -4 V
17 -4 V
18 -4 V
17 -4 V
18 -4 V
17 -4 V
17 -3 V
18 -4 V
17 -3 V
18 -3 V
17 -2 V
17 -2 V
18 -2 V
17 -1 V
18 0 V
17 -1 V
17 1 V
18 0 V
17 1 V
18 2 V
17 2 V
17 2 V
18 3 V
17 3 V
18 4 V
17 3 V
17 4 V
18 4 V
17 4 V
18 4 V
17 4 V
17 4 V
18 4 V
17 3 V
18 4 V
17 3 V
17 2 V
18 3 V
17 1 V
18 2 V
17 0 V
17 0 V
18 -1 V
17 -1 V
18 -2 V
17 -2 V
17 -3 V
18 -4 V
17 -4 V
18 -4 V
17 -4 V
17 -5 V
18 -4 V
17 -3 V
18 -3 V
17 -1 V
17 1 V
18 3 V
17 7 V
18 11 V
17 15 V
17 23 V
18 31 V
17 43 V
18 57 V
1.000 UL
LT2
1842 544 M
263 0 V
467 1373 M
18 -26 V
17 -19 V
18 -15 V
17 -10 V
17 -7 V
18 -5 V
17 -3 V
18 -2 V
17 0 V
17 0 V
18 1 V
17 2 V
18 1 V
17 2 V
17 2 V
18 2 V
17 2 V
18 2 V
17 1 V
17 1 V
18 1 V
17 1 V
18 0 V
17 0 V
17 -1 V
18 0 V
17 -1 V
18 -1 V
17 -1 V
17 -2 V
18 -1 V
17 -2 V
18 -2 V
17 -1 V
17 -2 V
18 -2 V
17 -2 V
18 -2 V
17 -2 V
17 -1 V
18 -2 V
17 -1 V
18 -2 V
17 -1 V
17 -1 V
18 0 V
17 -1 V
18 0 V
17 -1 V
17 1 V
18 0 V
17 1 V
18 0 V
17 1 V
17 1 V
18 2 V
17 1 V
18 2 V
17 1 V
17 2 V
18 2 V
17 2 V
18 2 V
17 2 V
17 1 V
18 2 V
17 2 V
18 1 V
17 2 V
17 1 V
18 1 V
17 1 V
18 0 V
17 1 V
17 0 V
18 0 V
17 -1 V
18 -1 V
17 -1 V
17 -1 V
18 -2 V
17 -2 V
18 -2 V
17 -2 V
17 -2 V
18 -1 V
17 -2 V
18 -1 V
17 0 V
17 0 V
18 2 V
17 3 V
18 5 V
17 7 V
17 10 V
18 15 V
17 19 V
18 26 V
1.000 UL
LT3
1842 444 M
263 0 V
467 1404 M
18 -4 V
17 -3 V
18 -3 V
17 -1 V
17 -1 V
18 -1 V
17 -1 V
18 0 V
17 0 V
17 0 V
18 0 V
17 1 V
18 0 V
17 0 V
17 0 V
18 1 V
17 0 V
18 0 V
17 0 V
17 1 V
18 0 V
17 0 V
18 0 V
17 0 V
17 0 V
18 0 V
17 0 V
18 -1 V
17 0 V
17 0 V
18 0 V
17 -1 V
18 0 V
17 0 V
17 0 V
18 -1 V
17 0 V
18 0 V
17 -1 V
17 0 V
18 0 V
17 0 V
18 -1 V
17 0 V
17 0 V
18 0 V
17 0 V
18 0 V
17 20 V
17 -20 V
18 0 V
17 0 V
18 0 V
17 0 V
17 0 V
18 1 V
17 0 V
18 0 V
17 0 V
17 1 V
18 0 V
17 0 V
18 1 V
17 0 V
17 0 V
18 0 V
17 1 V
18 0 V
17 0 V
17 0 V
18 1 V
17 0 V
18 0 V
17 0 V
17 0 V
18 0 V
17 0 V
18 0 V
17 -1 V
17 0 V
18 0 V
17 0 V
18 -1 V
17 0 V
17 0 V
18 0 V
17 -1 V
18 0 V
17 0 V
17 0 V
18 0 V
17 1 V
18 1 V
17 1 V
17 1 V
18 3 V
17 3 V
18 4 V
stroke
grestore
end
showpage
}}%
\put(1792,444){\makebox(0,0)[r]{$\zeta = 0.01$}}%
\put(1792,544){\makebox(0,0)[r]{$       \zeta = 0.05$}}%
\put(1792,644){\makebox(0,0)[r]{$\zeta = 0.1$}}%
\put(1792,744){\makebox(0,0)[r]{$\zeta = 0.2$}}%
\put(1320,50){\makebox(0,0){$X / \zeta$}}%
\put(100,910){%
\makebox(0,0)[b]{\shortstack{${\cal F}^S (X, \zeta) / [ {\cal F}^S (X, \zeta) - D (X/ \zeta)] $}}%
}%
\put(2190,200){\makebox(0,0){1}}%
\put(2016,200){\makebox(0,0){0.9}}%
\put(1842,200){\makebox(0,0){0.8}}%
\put(1668,200){\makebox(0,0){0.7}}%
\put(1494,200){\makebox(0,0){0.6}}%
\put(1320,200){\makebox(0,0){0.5}}%
\put(1146,200){\makebox(0,0){0.4}}%
\put(972,200){\makebox(0,0){0.3}}%
\put(798,200){\makebox(0,0){0.2}}%
\put(624,200){\makebox(0,0){0.1}}%
\put(450,200){\makebox(0,0){0}}%
\put(400,1409){\makebox(0,0)[r]{1}}%
\put(400,1224){\makebox(0,0)[r]{0.95}}%
\put(400,1039){\makebox(0,0)[r]{0.9}}%
\put(400,855){\makebox(0,0)[r]{0.85}}%
\put(400,670){\makebox(0,0)[r]{0.8}}%
\put(400,485){\makebox(0,0)[r]{0.75}}%
\put(400,300){\makebox(0,0)[r]{0.7}}%
\end{picture}%
\endgroup
\caption{The influence of including the D-term on the GRV98-model quark singlet GPD at the input scale in the ERBL region $X < \zeta$, for various values of $\zeta$. The anti-symmetry which the D-term shares with the singlet, about $X = \zeta/2$,   
is manifest by the symmetry about $X = \zeta/2$ in this ratio plot.} 
\label{fig:dterm} 
\end{figure}

Recall that in the definition of the unpolarized singlet-type quark input GPDs we included an extra piece in the ERBL region only, 
 known as the D-term, which we modelled following \cite{dterm} in eq.(\ref{eq:dterm}).  
This extra piece simply forms part of the definition of the input which is  
then evolved. In order to quantify its influence we consider in fig.(\ref{fig:dterm}) the ratio of the quark singlet 
with the D-term included to that with it omitted for various values of $\zeta$.  
One observes that its numerical influence becomes progressively less significant as $\zeta$ 
decreases (at $\zeta = 0.01$ it is less than $1\%$). 
 
\subsection{Influence of the profile functions}  

In section \ref{sec:inp} we specified in eq.(\ref{ddinp}) specific cases of Radyushkin model \cite{rad2,musrad} for GPDs based on double distributions. The original model specified a more general form for the profile functions:
\begin{equation}
\pi(x,y) = \frac{\Gamma(2b + 2)}{2^{2b+1} \Gamma^2 (b+1)} \frac{[(1 -|x|)^2 -y^2]^b}{(1 -|x|)^{2b+1}} \, . 
\label{profile}
\end{equation}
So that eq.(\ref{ddinp}) corresponds to $b=1$ for the quark and $b=2$ for the gluon. It was suggested in \cite{dterm} that variable $b$ could be used as a fit parameter to extract GPDs from DVCS observables. It was also noted that the limit of very large $b$ corresponds to the 
forward limit of $\zeta$-independent GPDs, for all $X > \zeta$. 

\begin{figure}  
\centering  
\mbox{\epsfig{file=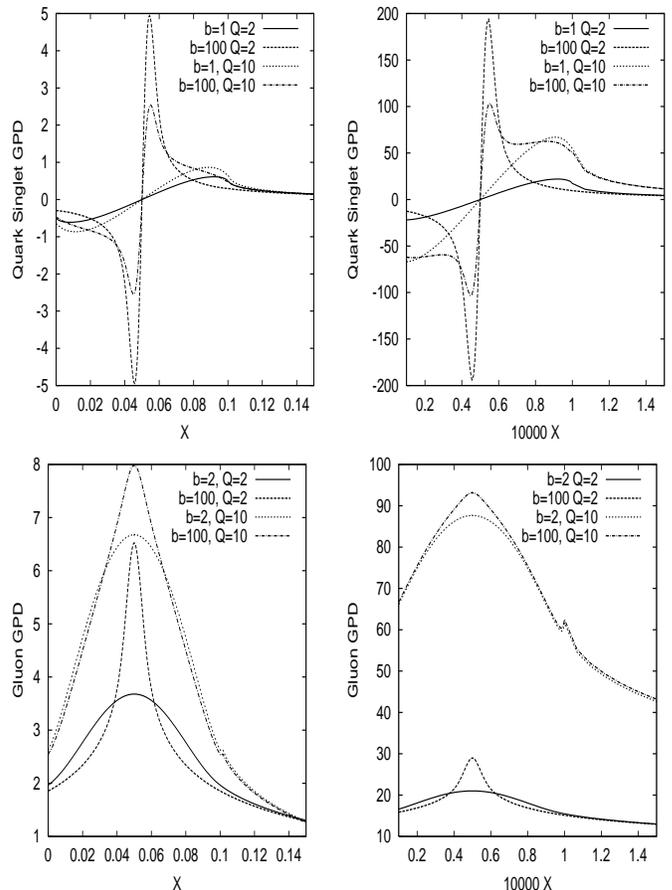,width=9.5cm,height=12cm}}  
\caption{Sensitivity of the input ($Q=2$~GeV) and evolved ($Q=10$~GeV) GPDs to the choice of the parameter 
$b$ for the profile functions in eq.(\ref{profile}). 
We compare large $b=100$ with the canonical values of eq.(\ref{ddinp}) ($b = 1.0$ for the quark singlet and $b=2$ for the gluon), 
using GRV98 forward distributions at both small $\zeta = 0.0001$ and large $\zeta =0.1$ skewedness (right and left hand plots 
respectively). The $b=100$ case is shown with thickened lines.}  
\label{fig:b100} 
\end{figure}

We have investigated this issue numerically and find that in the 
DGLAP region the GPDs tend to the PDFs very slowly as a function of $b$, particularly close to $X=\zeta$. Meanwhile the behaviour of the GPDs in the ERBL region is seen to change 
dramatically. We illustrate this in fig. \ref{fig:b100} at small and large $\zeta$ (i.e. $\zeta =0.0001, 0.1$, respectively),  
comparing $b=100$ with the canonical values for both the quark singlet and gluon, using GRV98 forward distributions. 
As can be seen, in the DGLAP region ($X > \zeta$) using such a large value of $b$ has only a marginal effect. 
In the ERBL region a rather dramatic effect is seen at the input scale, particularly close to the symmetric 
point $X=\zeta/2$. It is satisfying to observe (from the $Q=10$~GeV curves) that under evolution both choices appear 
to tend to similar evolved distributions (as the memory of the details of the input distribution are washed out by 
the skewed evolution). We may conclude from the figures that if parameter $b$ is to be used as a fit parameter to 
constrain GPDs using data, it would require high statistics data on DVCS observables which are very sensitive 
to the details of the real part of DVCS amplitudes in the ERBL region, such as the azimuthal angle asymmetry or the 
charge asymmetry (see e.g. \cite{bmns3}). We may also conclude that taking the very large $b$ limit of Radyushkin's 
ansatz is not a particularly good way of numerically probing models in which the GPD is the same as the PDF in the 
DGLAP region, since the convergence appears to be very slow.

\section{Conclusions}  
  
\label{sec:con}  
  
In this paper we have presented a complete numerical solution for the next-to-leading order  
(NLO) evolution equations for polarized and unpolarized generalised parton distributions (GPDs).  
We demonstrated that our solutions stay smooth and preserve all the symmetries of GPDs, giving us 
confidence in the correctness of our numerical implementation. We presented the relative effect  
of moving from leading-order (LO) to NLO for a common input.  
We illustrate the effect of NLO skewed evolution on the magnitude and shape of two  
correctly-symmetrized input models based on conventional parton density functions, for two  
values of the skewedness parameter, $\zeta  = 0.0001,0.1$, typical of HERA and HERMES kinematics. 
We demonstrate that in NLO, in contrast to LO, for small skewedness, the ratio of GPD gluon  
to PDF gluon in the unpolarized case does not have to be larger than unity in the 
$\overline{MS}$ scheme employed here. This new effect is directly relevant to the study of  
small-$x$ exclusive diffractive vector meson and real photon production at HERA. 
The next step is to use these NLO GPDs to make NLO predictions for measurable physical processes \cite{short,advcs,long}, such as DVCS, at HERA \cite{zeus,h1}, HERMES \cite{hermes} and Jefferson Lab \cite{jlab}. A detailed and direct comparison with all available data from a wide range of experimental observables is ultimately required to establish accurately the details of the GPDs.
 
\section*{Acknowledgements}  
  
\label{sec:ack}  
  
At the initial stages of this work, A.\ F.\ was supported by the E.\ U.\ contract $\#$ FMRX-CT98-0194, 
and then by the DFG under contract $\#$ FR 1524/1-1. M.\ M.\ was 
supported by PPARC.


\begin{thebibliography}{9}
\bibitem{foot1} A solution to the RGEs at NLO has been presented at large $x$ in \cite{bmns1,bmns2}, however due to technical problems the methods used could not be applied at smaller $x$.
\bibitem{foot2} The starting scale for GRV98 and GRSV00 are rather low. In order  
to make a meaningful comparison of the effects of skewedness, we evolve both polarized and  
unpolarized distributions to the common input scale of $Q_0 = 2$~GeV before introducing  
skewedness into the evolution. For computational reasons, we also found it convenient to use a  
faithful multi-parameter fit to these GRV and GRSV distribution rather than using the codes  
provided by the authors. For GRSV00 we choose the ``standard scenario'' option of unbroken polarised sea distributions.
\bibitem{foot3} The endpoint ($z=1$) concentrated terms in eqs.(\ref{dglapeq},\ref{erbleq}) can be found in \cite{bfm} and are determined by the requirement that the zeroth order conformal moment of the kernel has to be zero. Note that in the numerical implementation of the evolution the pure singlet component in the $qq$ kernel in both DGLAP and ERBL regions is not regularized (for more details see \cite{bfm}).
\bibitem{foot4} It was not applicable, even at LO, in the polarized case.
\bibitem{foot5} A slight deviation of the ratio from unity is observed in the plot for $X>0.1$  
because for GRV98, we use an analytic fit to the actual distributions at our input scale.  
This fit starts to deviate slightly from the results produced by the FORTRAN routines of GRV98 at 
the input scale. The NLO evolution enhances this difference and thus we 
do not expect the ratio to approach unity precisely for large $X \gg \zeta$. Note however 
that the deviation is at most $2\%$. The small wiggles in the curves are artifacts  
of the fitting procedure since we normalise to the code of GRV98.
\bibitem{foot6} Small deviations from unity are observed at very large
$X > 0.2$ for evolved scales. This results from the fact that in our
input GPDs we directly implement eqs.(9,10,13) and tables.(1,2) of
\cite{gehrst} (gluon A) analytically and the resultant numbers deviate
slightly from those obtained from the grid-based FORTRAN routine
provided by Gehrman-Stirling at such large $X$. Since we use the
latter to normalise this ratio we do not expect to reproduce the
forward limit with perfect accuracy.
\bibitem{nlolocomment} In the quark case, we used only the singlet
type combination of the down quark since it does not change sign as the up  
and the strange do. A change in sign distorts ratio plots by a large ratio around the point where the distributions changes sign, since this particular point is shifted towards lower $X$ in NLO compared to LO.
\end{thebibliography}
\end{document}